\documentclass[onecolumn,showpacs,amssymb,10pt]{revtex4}
\usepackage{epsfig}
\usepackage{bm}
\usepackage{color}

\def\ketm#1{  \left\vert  #1   \right\rangle   }

\def\sprm#1#2{  \left\langle #1 \left\vert \right. #2 \right\rangle   }

\def\mem#1#2#3{  \left\langle #1 \left\vert  #2 \right\vert #3 \right\rangle   }
\def\rmem#1#2#3{  \left\langle #1 \left\vert \left\vert  #2
                  \right\vert \right\vert #3 \right\rangle   }

\def\sixjm#1#2#3#4#5#6{  \left\{ \begin{array}{ccc}
                                               #1 & #2 & #3  \\
                                               #4 & #5 & #6
                     \end{array} \right\}   }

\begin{document}

%
%

\title{Angular correlations in the two--photon decay of helium--like heavy ions}

%
%

\author{A.~Surzhykov$^{1,2}$, A.~Volotka$^{3,4}$,
F.~Fratini$^{1,2}$, J.~P.~Santos$^{5}$, P.~Indelicato$^{6}$,
G.~Plunien$^{3}$, Th.~St\"ohlker$^{1,2}$, S.~Fritzsche$^{2,7}$}

\affiliation{{}\\
$^{1}$Physikalisches Institut,
Universit\"at Heidelberg, D--69120 Heidelberg, Germany \\
$^{2}$GSI Helmholtzzentrum f\"ur Schwerionenforschung, D--64291 Darmstadt, Germany \\
$^{3}$Institut f\"ur Theoretische Physik, Technische Universit\"at Dresden, D--01062 Dresden, Germany\\
$^{4}$Department of Physics, St. Petersburg State University,
Petrodvorets, 198504 St. Petersburg, Russia \\
$^{5}$Centro de F\'isica At\'omica, Departamento de F\'isica,
Faculdade de Ci\^encias e Tecnologia,
FCT, Universidade Nova de Lisboa, 2829-516 Caparica, Portugal\\
$^{6}$Laboratoire Kastler Brossel, \'Ecole Normale Sup\'erieure,
CNRS, Universit\'e P. et M. Curie -- Paris 6, Case 74; 4, place
Jussieu, 75252 Paris CEDEX 05, France\\
$^{7}$Department of Physical Sciences, P.O. Box 3000, Fin--90014
University of Oulu, Finland}

\date{\today}

%
%
%
%

\begin{abstract}
The two--photon decay of heavy, helium--like ions is investigated
based on second--order perturbation theory and Dirac's relativistic
equation. Special attention has been paid to the angular emission of
the two photons, i.e., how the angular correlation function depends
on the shell structure of the ions in their initial and final
states. Moreover, the effects from the (electric and magnetic)
non-dipole terms in the expansion of the electron--photon
interaction are discussed. Detailed calculations have been carried
out for the two--photon decay of the $1s2s \, ^1S_0$, $1s2s \,
^3S_1$ and $1s2p \, ^3P_0$ states of helium--like xenon Xe$^{52+}$,
gold Au$^{77+}$ and uranium U$^{90+}$ ions.
\end{abstract}


\maketitle

%
%
%
%

%
%
\section{Introduction}

Owing to the recent advances in heavy-ion accelerator and trap
facilities as well as in detection techniques, new possibilities
arise to study the \textit{electronic structure} of simple atomic
systems in strong (nuclear) Coulomb fields. Relativistic, quantum
electrodynamics (QED), or even parity non--conservation (PNC)
effects, which are difficult to isolate in neutral atoms, often
become enhanced in high--$Z$, few--electron ions. In order to
improve our understanding of these fundamental interactions, a
number of experiments have been recently carried out on the
characteristic photon emission from heavy ions
\cite{FrI05,GuS05,TrK09}. Apart from the one--photon bound--bound
transitions, the two--photon decays of metastable ionic states have
also attracted much interest since the analysis of its properties
may reveal unique information about the \textit{complete} spectrum
of the ion, including negative energy (positron) solutions of
Dirac's equation. Until now, however, most two--photon studies were
focused on measuring the total and energy--differential decay rates
\cite{DuB93,AlA97,ScM99,MoD04,IlU06,KuT09,TrS09} which were found in
good agreement with theoretical predictions, based on relativistic
calculations
\cite{DrG81,GoD81,Dra86,ToL90,DeJ97,SaP98,KoF03,LaS04,LaS05,SuS09}.
In contrast, much less attention has been previously paid to the
angular and polarization correlations between the emitted photons.
The first two--photon correlation studies with heavy ions are likely
to be carried out at the GSI facility in Darmstadt, where the
significant progress has been recently made in development of solid
state, position sensitive x--ray detectors \cite{TaS06,StS07}. By
means of these detectors, a detailed analysis of the angular and
polarization properties of two--photon emission will become possible
and will provide more insights into the electronic structure of
heavy, few--electron ions.

\medskip

Despite of the recent interest in two--photon coincidence studies,
not much theoretical work has been done so far to explore the
photon--photon correlations in the decay of heavy atomic systems.
While some predictions are available for the hydrogen--like
\cite{Au76,SuK05} and neutral atoms \cite{ToL90}, no systematic
angular (and polarization) analysis was performed for the
\textit{helium--like} ions which are the most suitable candidates
for two--photon investigations in high--$Z$ domain. In the present
work, therefore, we apply the second--order perturbation theory
based on Dirac's equation to investigate the $\gamma - \gamma$
angular correlations in the decay of two--electron systems. The
basic relations of such a (relativistic) perturbation approach will
be summarized in the following Section. In particular, here we
introduce the second--order transition amplitude that describes a
bound--state transition under the simultaneous emission of two
photons. The evaluation of these (many--body) amplitudes within the
framework of the independent particle model (IPM) is thereafter
discussed in Section \ref{sect_ipm}. Within this approximation that
is particularly justified for the high--$Z$ regime
\cite{Dun96,Dra89,Dun04}, the photon--photon correlation function
from Section \ref{sect_angular_correlation} can be traced back to
the one--electron matrix elements. This reduction enables us to
implement the well--established Green's function as well as finite
basis set methods \cite{Ind95,SaJ96,ShT04} and to calculate the
correlation functions for the $1s 2s \, ^1S_0 \to 1s^2 \, ^1S_0$,
$1s 2s \, ^3S_1 \to 1s^2 \, ^1S_0$ and $1s 2p \, ^3P_0 \to 1s^2 \,
^1S_0$ transitions in helium--like xenon Xe$^{52+}$, gold Au$^{77+}$
and uranium U$^{90+}$ ions. Results from these computations are
displayed in Section \ref{sect_results} and indicate a strong
dependence of the photon emission pattern on the symmetry and parity
of initial and final ionic states. Moreover, the significant effects
that arise due the higher--multipole terms in the expansion of the
electron--photon interaction are also discussed in the context of
angular correlation studies. Finally, a brief summary is given in
Section \ref{sect_summary}.

%
%
\section{Theoretical background}

\subsection{Second--order transition amplitude}

Since the second--order perturbation theory has been frequently
applied in studying two--photon decay, here we may restrict
ourselves to a short compilation of the basic formulas relevant for
our analysis and refer for all further details to the literature
\cite{DrG81,GoD81,ToL90,SaP98,LaS04,LaS05,SuK05,Dra89,Dun04}. Within
the \textit{relativistic} framework, the second--order transition
amplitude for the emission of two photons with wave vectors
$\bm{k}_i$ ($i$ = 1, 2) and polarization vectors
$\bm{u}_{\lambda_i}$ ($\lambda = \pm 1$ ) is given by
\begin{widetext}
\begin{eqnarray}
   \label{matrix_general}
   {\cal M}_{fi}(M_f, M_i, \lambda_1, \lambda_2) &=& \sum\limits_{\gamma_\nu J_\nu M_\nu}
   \frac{
   \mem{\gamma_f J_f M_f}{\hat{\mathcal{R}}^\dag(\bm{k}_1, \bm{u}_{\lambda_1})}{\gamma_\nu J_\nu M_\nu}
   \mem{\gamma_\nu J_\nu M_\nu}{\hat{\mathcal{R}}^\dag(\bm{k}_2, \bm{u}_{\lambda_2})}{\gamma_i J_i M_i}
   }{E_\nu - E_i + \omega_2} \nonumber \\
   &+&
   \sum\limits_{\gamma_\nu J_\nu M_\nu}
   \frac{
   \mem{\gamma_f J_f M_f}{\hat{\mathcal{R}}^\dag(\bm{k}_2, \bm{u}_{\lambda_2})}{\gamma_\nu J_\nu M_\nu}
   \mem{\gamma_\nu J_\nu M_\nu}{\hat{\mathcal{R}}^\dag(\bm{k}_1, \bm{u}_{\lambda_1})}{\gamma_i J_i M_i}
   }{E_\nu - E_i + \omega_1} \, ,
\end{eqnarray}
\end{widetext}
where $\ketm{\gamma_i J_i M_i}$ and $\ketm{\gamma_f J_f M_f}$ denote
the (many--electron) states with well--defined total angular momenta
$J_{i,f}$ and their projections $M_{i,f}$ of the ions just before
and after their decay, and $\gamma_{i,f}$ all the additional quantum
numbers as necessary for a unique specification. The energies of
these states, $E_i$ and $E_f$, are related to the energies
$\omega_{1,2} = c k_{1,2}$ of the emitted photons by the energy
conservation condition:
\begin{equation}
   \label{energy_conservation}
   E_i - E_f = \hbar \omega_1 + \hbar \omega_2 \, .
\end{equation}
Using this relation, it is convenient to define the so--called
energy sharing parameter $y = \omega_1/(\omega_1 + \omega_2)$, i.e.,
the fraction of the energy which is carried away by the first
photon.

\medskip

In Eq.~(\ref{matrix_general}), moreover, $\hat{\mathcal{R}}$ is the
transition operator that describes the interaction of the electrons
with the electromagnetic radiation. In velocity (Coulomb) gauge for
the coupling of the radiation field this operator can be written as
a sum of one--particle operators:
\begin{equation}
   \label{A_operator_general}
   \hat{\mathcal{R}}(\bm{k}, \bm{u}_{\lambda}) =
   \sum_{m} {\bm \alpha_m} \, \mathcal{A}_{\lambda, m} =
   \sum_{m} {\bm \alpha_m} \,  {\bm u}_{\lambda} \,
   {\rm e}^{i {\bm k} \cdot {\bm r}_m}  \, ,
\end{equation}
where ${\bm \alpha}_m = \left( \alpha_{x,m}, \alpha_{y,m},
\alpha_{z,m} \right)$ denotes the vector of the Dirac matrices for
the $m$--th particle and $\mathcal{A}_{\lambda,m}$ the vector
potential of the radiation field. To further simplify the
second--order transition amplitude (\ref{matrix_general}) for
practical computations, it is convenient to decompose the potential
$\mathcal{A}_{\lambda,m}$ into spherical tensors, i.e. into its
electric and magnetic \textit{multipole} components. For the
emission of the photon in the direction $\hat{\bm k} = (\theta,
\phi)$ with respect to the quantization ($z$--) axis such a
decomposition reads \cite{Ros53}:
\begin{eqnarray}
   \label{multipole_expansion}
   \bm{u}_{\lambda} {\rm e}^{i \bm{k} \cdot \bm{r}} =
   \sqrt{2 \pi} \sum\limits_{L =1}^{\infty} \, \sum\limits_{M=-L}^{L}
   \, \sum\limits_{p \, = 0, 1}i^L \, [L]^{1/2} \,
   (i \lambda)^p \, \hat{a}^{p}_{LM}(k) \, D^L_{M \lambda}(\hat{\bm k})
   \, ,
\end{eqnarray}
where $[L] \equiv 2L + 1$, $D^L_{M \lambda}$ is the Wigner rotation
matrix of rank $L$ and $\hat{a}^{p \, = 0,1}_{LM}(k)$ refer to
magnetic ($p$=0) and electric ($p$=1) multipoles, respectively.

\medskip

The multipole decomposition of the photon field in terms of its
irreducible components with well--defined transformation properties
enables us to simplify the second--order amplitude by employing the
techniques from Racah's algebra. Inserting
Eqs.~(\ref{A_operator_general})--(\ref{multipole_expansion}) into
the matrix element (\ref{matrix_general}) and by making use of the
Wigner--Eckart theorem, we obtain:
\begin{eqnarray}
   \label{M_amplitude_new}
   {\cal M}_{fi}(M_f, M_i, \lambda_1, \lambda_2) &=& 2 \pi \,
   \sum\limits_{L_1 M_1 p_1} \sum\limits_{L_2 M_2 p_2}
   (-i)^{L_1 + L_2} \, [L_1, L_2]^{1/2} \, (-i \lambda_1)^{p_1} \,
   (-i \lambda_2)^{p_2} \, D^{L_1 *}_{M_1 \lambda_1}(\hat{k}_1) \,
   D^{L_2 *}_{M_2 \lambda_2}(\hat{k}_2) \nonumber \\
   &\times& \sum\limits_{J_\nu M_\nu}
   \frac{1}{[J_i, J_\nu]^{1/2}}
   \Bigg[
   \sprm{J_f M_f \, L_1 M_1}{J_\nu M_\nu} \, \sprm{J_\nu M_\nu \, L_2 M_2}{J_i
   M_i} \,
   S_{L_1 p_1, \, L_2 p_2}^{J_\nu}(\omega_2) \nonumber \\
   &+&
   \sprm{J_f M_f \, L_2 M_2}{J_\nu M_\nu} \, \sprm{J_\nu M_\nu \, L_1 M_1}{J_i M_i} \,
   S_{L_2 p_2, \, L_1 p_1}^{J_\nu}(\omega_1) \Bigg] \,
   ,
\end{eqnarray}
where the second--order \textit{reduced} transition amplitude is
given by:
\begin{eqnarray}
   \label{S_function}
   S^{J_\nu}_{L_1 p_1, \, L_2 p_2}(\omega_2)
   &=&
   \sum\limits_{\gamma_\nu} \frac{\rmem{\gamma_f J_f}{\sum\limits_{m} \bm{\alpha}_m  \,
   \hat{a}^{p_1 \dag}_{L_1, m}(k_1)}{\gamma_\nu J_\nu}
   \rmem{\gamma_\nu J_\nu}{\sum\limits_{m} \bm{\alpha}_m \,
   \hat{a}^{p_2 \dag}_{L_2, m}(k_2)}{\gamma_i J_i}
   }{E_\nu - E_i + \omega_2} \, .
\end{eqnarray}
Here, the summation over the intermediate states formally runs over
the complete spectrum of the ions, including a summation over the
discrete part of the spectrum as well as the integration over the
positive-- and negative--energy continua. In practice, such a
summation is not a simple task especially when performed over the
\textit{many--electron} states $\ketm{\gamma_\nu J_\nu}$. In the
next section, therefore, we shall employ the independent particle
model in order to express the reduced matrix elements
(\ref{S_function}) for many--electron ions in terms of their
one--electron analogs.

\subsection{Evaluation of the reduced transition amplitudes}
\label{sect_ipm}

As seen from Eqs.~(\ref{M_amplitude_new})--(\ref{S_function}), one
has first to generate a \textit{complete} set of many--electron
states $\ketm{\gamma J}$ in order to calculate the second order
transition amplitude $M_{fi}$. A number of approximate methods, such
as multi--configuration Dirac--Fock (MCDF) \cite{Gra89,Fri02} and
configuration interaction (CI) \cite{DeJ97}, are known in atomic
structure theory for constructing these states. Moreover, the
systematic perturbative QED approach in combination with the CI
method turned out to be most appropriate for describing both
transition probabilities \cite{TuV05,InS04} and transition energies
\cite{ArS05} in highly charged ions. In the high--$Z$ domain,
however, the structure of few--electron ions can already be
reasonably well understood within the independent particle model
(IPM). This model is well justified for heavy species especially,
since the interelectronic effects scale with $1/Z$ and, hence, are
much weaker than the electron--nucleus interaction
\cite{Dra89,Dun04,SuJ08}. Within the IPM, that takes the Pauli
principle into account, the many--electron wave functions are
approximated by means of Slater determinants, built from
one--particle orbitals. For this particular choice of the
many-electron function, all the (first-- and the second--order)
matrix elements can be easily decomposed into the corresponding
single--electron amplitudes.

\medskip

For a helium--like system, the decomposition of the reduced
amplitude (\ref{S_function}) reads:
\begin{eqnarray}
   \label{S_function_decomposition}
   S^{J_\nu}_{L_1 p_1, \, L_2 p_2}(\omega_2)
   &=& -\delta_{J_\nu L_1} \, [J_i, J_\nu]^{1/2}
   \sum\limits_{j_\nu} (-1)^{J_i + J_\nu + L_2}
   \, \sixjm{j_\nu}{j_0}{J_\nu}{J_i}{L_2}{j_i}
   \, S^{j_\nu}_{L_1 p_1, \, L_2 p_2}(\omega_2) \, ,
\end{eqnarray}
where the \textit{one--electron} matrix elements of the (electric
and magnetic) multipole field operators are given by
\begin{eqnarray}
   \label{S_one_electron}
   S^{j_\nu}_{L_1 p_1, \, L_2 p_2}(\omega_2) =
   \sum\limits_{n_\nu}
   \frac{\rmem{n_0 j_0}{\bm{\alpha}  \,
   \hat{a}^{p_1 \dag}_{L_1}(k_1)}{n_\nu j_\nu}
   \rmem{n_\nu j_\nu}{\bm{\alpha}  \, \hat{a}^{p_2 \dag}_{L_2}(k_2)}{n_i j_i}
   }{E_\nu - E_i + \omega_2} \, .
\end{eqnarray}
We assume here that the ``spectator'' electron, being in hydrogenic
state $\ketm{n_0 j_0}$, stays passive in the decay process.
Moreover, $\ketm{n_i j_i}$, $\ketm{n_\nu j_\nu}$ and $\ketm{n_f j_f}
= \ketm{n_0 j_0}$ denote the initial, intermediate and final states
of the ``active'' electron, correspondingly. The great advantage of
formula (\ref{S_function_decomposition}) is that it helps us to
immediately evaluate the many--electron transition amplitude
(\ref{S_function}) in terms of the (one--particle) functions
$S^{j_\nu}_{L_1 p_1, \, L_2 p_2}(\omega_2)$. The summation over the
\textit{complete} one--particle spectrum that occurs in these
functions can be performed by means of various methods. In the
present work, we make use of (i) the relativistic Coulomb--Green's
function \cite{SuK05,JeS08,SuKCPC05} and (ii) a B--spline discrete
basis set \cite{SaP98,Ind95,SaJ96,ShT04,SuS09} to evaluate all the
second--order transition amplitudes. Indeed, both approaches yield
almost identical results for the angular correlation functions in
the two--photon decay of heavy helium--like ions.

\begin{figure}
\includegraphics[height=8.5cm,angle=00,clip=]{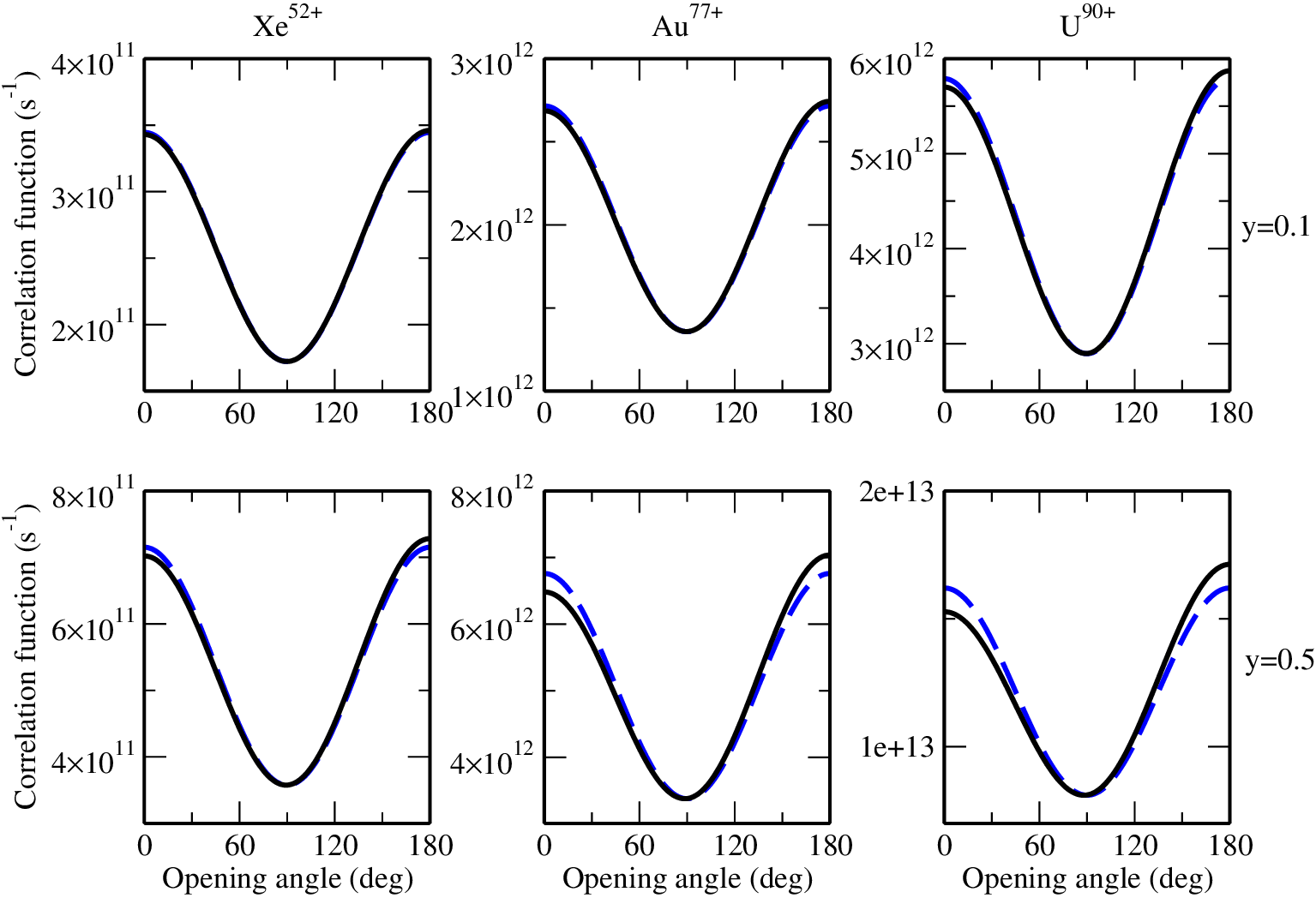}
\vspace*{0.0cm} \caption{(Color online) Angular correlation function
(\ref{function_definition}) for the $1s 2s \, ^1S_0 \to 1s^2 \,
^1S_0$ two--photon decay of helium--like xenon, gold and uranium
ions. Calculations obtained within the electric dipole 2E1
approximation (dashed line) are compared with those including all
the allowed multipoles (solid line). Results are presented for the
relative photon energies $y$ = 0.1 (upper panel) and 0.5 (lower
panel).} \label{Fig1}
\end{figure}
\subsection{Differential decay rate}
\label{sect_angular_correlation}

Equation (\ref{M_amplitude_new}) displays the general form of the
relativistic transition amplitude for the two--photon decay of
many--electron ions. Such an amplitude represents the ``building
block'' for studying various properties of the emitted radiation.
For instance, the differential two--photon decay rate can be written
in terms of (squared) transition amplitudes as:
\begin{equation}
   \label{diff_rate}
   \frac{{\rm d}w}{{\rm d}\omega_1 {\rm d}\Omega_1 {\rm d}\Omega_2} =
   \frac{\omega_1 \omega_2}{(2 \pi)^3 c^3} \,
   \frac{1}{2J_i + 1} \, \sum\limits_{M_i M_f} \, \sum\limits_{\lambda_1 \lambda_2}
   \left|{\cal M}_{fi}(M_f, M_i, \lambda_1, \lambda_2) \right|^2 \, ,
\end{equation}
if we assume that the excited ions are initially unpolarized and
that the spin states of the emitted photons remain unobserved in a
particular measurement. As seen from expression (\ref{diff_rate}),
the two--photon rate is \textit{single} differential--owing to the
conservation law (\ref{energy_conservation})---in the energy of one
of the photons but \textit{double} differential in the emission
angles. Accordingly, its further evaluation requires to determine
the \textit{geometry} under which the photon emission is considered.
Since no particular direction is preferred for the decay of an
unpolarized (as well as unaligned) ion, it is convenient to adopt
the quantization ($z$--) axis along the momentum of the ``first''
photon: $\bm{k}_1 || z$. Such a choice of the quantization axis
allows us to simplify the rate (\ref{diff_rate}) and to define the
\textit{angular correlation function}:
\begin{equation}
   \label{function_definition}
   W_{2\gamma}(\theta, y) = 8 \pi^2 \, (E_i - E_f) \,
   \frac{{\rm d}w}{{\rm d}\omega_1 {\rm d}\Omega_1 {\rm d}\Omega_2} (\theta_1 = 0, \phi_1 = 0, \phi_2 = 0)
   \, ,
\end{equation}
that is characterized (apart from the relative energy $y$) by the
single polar angle $\theta = \theta_2$ of the ``second'' photon
momentum with respect to this axis. In this expression, moreover,
the factor $8\pi^2$ arises from the integration over the solid angle
$d\Omega_1 = \sin\theta_1 {\rm d}\theta_1 {\rm d}\phi_1$ of the
first photon as well as the integration over the azimuthal angle
${\rm d}\phi_2$ of the second photon. In the next Section, we shall
investigate the dependence of the function $W_{2\gamma}(\theta, y)$
on this \textit{opening angle} $\theta$ for various bound--bound
transitions and for a range of (relative) photon energies.

%
%
\section{Results and discussion}
\label{sect_results}

With the formalism developed above, we are ready now to analyze the
angular correlations in the two--photon decay of helium--like heavy
ions. In nowadays experiments, the excited states of these ions can
be efficiently populated in relativistic ion--atom collisions. For
example, the formation of the metastable $1s 2s \, ^1S_0$ state
during the inner--shell impact ionization of (initially)
lithium--like heavy ions has been studied recently at the GSI
storage ring in Darmstadt \cite{RzS06}. The radiative deexcitation
of this state can proceed only via the two--photon transition $1s 2s
\, ^1S_0 \to 1s^2 \, ^1S_0$ since a single--photon decay to the
$1s^2\;\, ^1S_0$ ground state is strictly forbidden by the
conservation of angular momentum. Fig.~1 displays the photon--photon
angular correlation function for this experimentally easily
accessible decay of helium--like xenon Xe$^{52+}$, gold Au$^{77+}$
and uranium U$^{90+}$ ions and for the two energy sharing parameters
$y$ = 0.1 (upper panel) and $y$ = 0.5 (lower panel). Moreover,
because the radiative transitions in high--$Z$ ions are known to be
affected by the higher terms of the electron--photon interaction
(\ref{A_operator_general}), calculations were performed within both,
the exact relativistic theory (solid line) to include all allowed
multipole components ($p_1 L_1, \, p_2 L_2$) in the amplitude
(\ref{M_amplitude_new}) as well as the electric dipole approximation
(dashed line), if only a single term with $L_1 = L_2 = 1$ and $p_1 =
p_2 = 1$ is taken into account. In the dipole 2E1 approach, as
expected, the angular distribution is well described by the formula
$1 + \cos^2\theta$ that predicts a \textit{symmetric}---with respect
to the opening angle $\theta = 90^\circ$---emission pattern of two
photons. Within the exact relativistic theory, in contrast, an
asymmetric shift in the angular correlation function is obtained. As
can be deduced from
Eqs.~(\ref{S_function_decomposition})--(\ref{function_definition}),
this shift arises from the interference between the leading 2E1
decay channel and higher multipole terms in the electron--photon
interaction:
\begin{equation}
   \label{W_1}
   W_{2\gamma}(\theta, y) \propto (1 + \cos^2\theta)
   + 4 \, \frac{\mathcal{S}_{M1}}{\mathcal{S}_{E1}} \, \cos\theta
   - \frac{20}{3} \, \frac{\mathcal{S}_{E2}}{\mathcal{S}_{E1}} \, \cos^3\theta
   + ... \, ,
\end{equation}
where, for the sake of brevity, we have introduced the notation
$\mathcal{S}_{Lp} = S^{J_\nu = L}_{L p, \, L p}(\omega_1) + S^{J_\nu
=L}_{L p, \, L p}(\omega_2)$. For high--$Z$ domain, the photon
emission occurs predominantly in the backward directions if the
nondipole terms are taken into account; an effect which becomes more
pronounced for the equal energy sharing (cf. bottom panel of
Fig.~1). Including the higher multipoles into the photon--photon
correlation function, a similar asymmetry was found in the past for
the $2s_{2/1} \to 1s_{1/2}$ decay in hydrogen--like heavy ions both
within the nonrelativistic \cite{Au76} and relativistic \cite{SuK05}
theory.

\medskip

\begin{figure}
\includegraphics[height=8.5cm,angle=0,clip=]{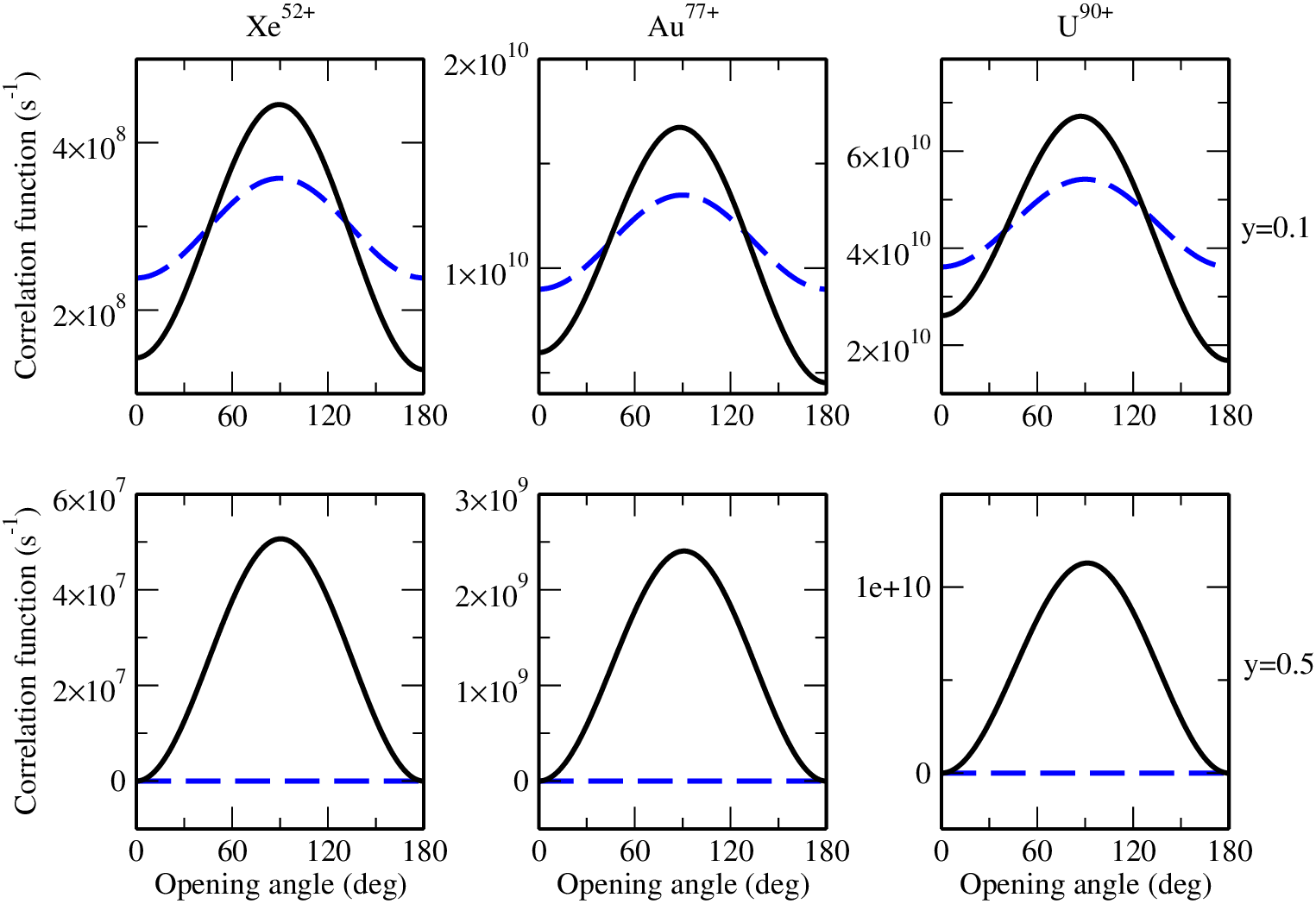}
\vspace*{0.0cm} \caption{(Color online) Angular correlation function
(\ref{function_definition}) for the $1s 2s \, ^3S_1 \to 1s^2 \,
^1S_0$ two--photon decay of helium--like xenon, gold and uranium
ions. Calculations obtained within the electric dipole 2E1
approximation (dashed line) are compared with those including all of
the allowed multipoles (solid line). Results are presented for the
relative photon energies $y$ = 0.1 (upper panel) and 0.5 (lower
panel).} \label{Fig2}
\end{figure}

Apart from the singlet $1s 2s \, ^1S_0$, the formation of the
triplet $1s 2s \, ^3S_1$ state has been also observed in recent
collision experiments at the GSI storage ring \cite{RzS06,KuT09}.
Although much weaker in intensity (owing to the dominant M1
transition), the two--photon decay of this $1s2s \, ^3S_1$ state has
attracted recent interest and might provide an important testing
ground for symmetry violations of Bose particles \cite{Dun04,DeB99}.
The angular correlation between the photons emitted in this $1s 2s
\, ^3S_1 \to 1s^2 \, ^1S_0$ (two--photon) decay is displayed in
Fig.~2, by comparing again the results from the exact relativistic
theory with the 2E1 dipole approximation. As seen from the figure,
the photon--photon correlation functions for the $2 ^3S_1 \to 1
^1S_0$ transition is much more sensitive with regard to higher
multipoles in the electron--photon interaction than obtained for the
$2 ^1S_0 \to 1 ^1S_0$ decay. The strongest non--dipole effect can be
observed for the equal energy sharing ($y$ = 0.5), where the
two--photon emission is strictly \textit{forbidden} within the
electric dipole approximation. This suppression of the 2E1 decay is
a direct consequence of the exchange symmetry of photons as required
by the Bose--Einstein statistics and, hence, a particular case of
the Landau--Yang theorem that forbids the decay of vector particles
into two photons (cf. Refs.~\cite{Dun04,DeB99,Lan48,Yan50} for
further details). In contrast to the 2E1 channel, the E1M2 $2 ^3S_1
\to 1 ^1S_0$ transition can proceed even if the energies of the two
photons are equal. This transition as well as higher multipole terms
give rise to a strongly anisotropic correlation function that
vanishes for the parallel and back--to--back photon emission and has
a maximum at $\theta = 90^\circ$.

\medskip

Large effects due to the higher multipole contributions to the $1s
2s \, ^3S_1 \to 1s^2 \, ^1S_0$ two--photon transition can be
observed not only for the case of equal energy sharing ($y$ = 0.5).
For the relative energy $y$ = 0.1, for example, the photon--photon
angular correlation function is found symmetric with regard to
$\theta = 90^\circ$ in the electric dipole (2E1) approximation but
becomes asymmetric in an exact relativistic theory. In contrast to
the decay of $2 ^1S_0$ state, however, a predominant
\textit{parallel emission} of both photons appears to be more likely
if the higher multipoles are taken into account. For the $2 ^3S_1
\to 1 ^1S_0$ two--photon decay of helium--like uranium U$^{90+}$,
for example, the intensity ratio $W_{2\gamma}(\theta = 0^\circ,
y=0.1)/W_{2\gamma}(\theta = 180^\circ, y=0.1)$ increases from
\textit{unity} within the electric dipole approximation to almost
1.6 in the exact relativistic treatment.

\medskip

Until now we have discussed the photon--photon correlations in the
decay of $1s 2s$ (singlet and triplet) helium--like states. Besides
these---well studied---transitions, recent theoretical interest has
been focused also on the $1s 2p \, ^3P_0 \to 1s^2 \, ^1S_0$
two--photon decay whose properties are expected to be sensitive to
(parity violating) PNC phenomena in atomic systems \cite{Dun96}.
Future investigations on such subtle parity non--conservation
effects will require first detailed knowledge on the angle (and
polarization) properties of two--photon emission as well as the role
of non--dipole contributions. The angular correlation function
(\ref{function_definition}) for the $2 ^3P_0 \to 1 ^1S_0$ transition
is displayed in Fig.~3, again, for two relative photon energies $y$
= 0.1 and 0.5 and for the nuclear charges $Z$ = 54, 79 and 92.
Calculations have been performed both within the exact theory and
the (``electric and magnetic'') dipole approximation which accounts
for the leading E1M1--M1E1 decay channel. As seen from the figure,
the emission pattern strongly depends on the energy sharing between
the photons. If, for example, one of the photons is more energetic
than the second one their parallel emission becomes dominant (cf.
upper panel of Fig.~3). In contrast, photons with equal energies,
i.e. when $y$ = 0.5, are more likely to be emitted back--to--back
while the differential rate (\ref{diff_rate}) vanishes identically
for $\theta = 0^\circ$. Such a behaviour of the photon--photon
angular correlation function is caused by the interference between
\textit{two} pathways which appear for each multipole component of
the $2 ^3P_0 \to 1 ^1S_0$ transition. For instance, the leading
E1M1--M1E1 decay may proceed either via intermediate $n \, ^3S_1$ or
$n \, ^3P_1$ states, thus given rise to a ``double--slit'' picture
that becomes most pronounced for the equal energy sharing. Simple
analytical expression for the angular correlation function which
accounts for such a Young--type interference effect can be obtained
from Eqs.~(\ref{S_function_decomposition})--(\ref{diff_rate}) as:
\begin{eqnarray}
   \label{W_2}
   W_{2\gamma}(\theta, y) &\propto& \sin^4\theta/2 \, \left| \mathcal{S}_{E1M1}
   \right|^2 \,
   \left( 1+ 2(1 + 2 \cos\theta) \frac{\mathcal{S}_{E2M2}}{\mathcal{S}_{E1M1}}
   \right) \nonumber \\
   &+& \cos^4\theta/2 \, \left| \mathcal{D}_{E1M1}
   \right|^2 \,
   \left( 1 - 2(1 - 2 \cos\theta) \frac{\mathcal{D}_{E2M2}}{\mathcal{D}_{E1M1}}
   \right) \, + ... ,
\end{eqnarray}
where, similar as before, we denote $\mathcal{S}_{L p_1, L p_2} =
S^{J_\nu = L}_{L p_1, \, L p_2}(\omega_1) + S^{J_\nu =L}_{L p_1, \,
L p_2}(\omega_2) + S^{J_\nu = L}_{L p_2, \, L p_1}(\omega_2) +
S^{J_\nu =L}_{L p_2, \, L p_1}(\omega_1)$ and $\mathcal{D}_{L p_1, L
p_2} = S^{J_\nu = L}_{L p_1, \, L p_2}(\omega_1) - S^{J_\nu =L}_{L
p_1, \, L p_2}(\omega_2) + S^{J_\nu = L}_{L p_2, \, L p_1}(\omega_2)
- S^{J_\nu =L}_{L p_2, \, L p_1}(\omega_1)$. Obviously, if the
energies of the two photons are equal, $\omega_1 = \omega_2$, the
second term in Eq.~(\ref{W_2}) turns to be \textit{zero} and the
photon emission is described by the $\sin^4 \theta/2$ angular
distribution modified by the non--dipole terms in the expansion of
electron--photon interaction. As seen from the lower panel of
Fig.~3, the contribution from these terms becomes more pronounced
for the back--to--back photon emission ($\theta$ = 180$^\circ$)
where they lead to about a 30 \% enhancement of the correlation
function. It is interesting to note that such an enhancement remains
almost constant along the helium isoelectronic sequence for $Z \ge$
54 due to similar ($\propto Z^{12}$) scaling of the E1M1 and E2M2
transition probabilities. Therefore, our calculations clearly
indicate the importance of higher multipoles for analyzing the
photon--photon correlations not only for high--$Z$ domain but also
for medium--$Z$ ions.

\begin{figure}
\includegraphics[height=8.5cm,angle=0,clip=]{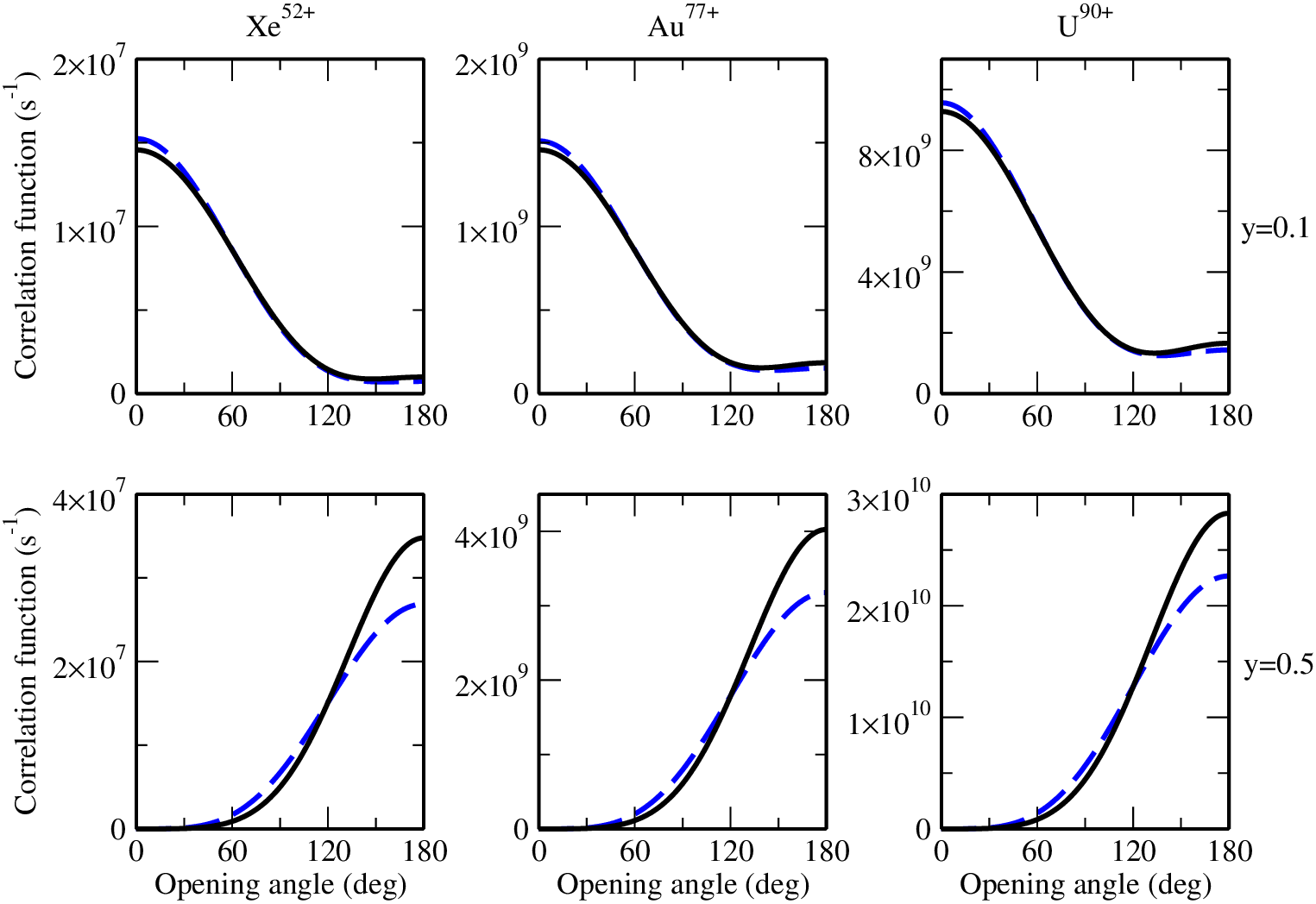}
\vspace*{0.0cm} \caption{(Color online) Angular correlation function
(\ref{function_definition}) for the $1s 2p \, ^3P_0 \to 1s^2 \,
^1S_0$ two--photon decay of helium--like xenon, gold and uranium
ions. Calculations obtained within the dipole E1M1 approximation
(dashed line) are compared with those including all of the allowed
multipoles (solid line). Results are presented for the relative
photon energies $y$ = 0.1 (upper panel) and 0.5 (lower panel).}
\label{Fig3}
\end{figure}
%
%
%

%
%
\section{Summary and outlook}
\label{sect_summary}

In summary, the two--photon decay of heavy, helium--like ions has
been investigated within the framework of the relativistic
second--order perturbation theory and the independent particle
model. In this study, special emphasis was placed on the angular
correlations between the emitted photons. A general expression for
the photon--photon correlation function was derived that accounts
for the complete expansion of the radiation field in terms of its
multipole components. Based on solutions of Dirac's equation, this
function has been calculated for the two--photon decay of the $1s2s
\, ^1S_0$, $1s2s \, ^3S_1$ and $1s2p \, ^3P_0$ states of
helium--like xenon Xe$^{52+}$, gold Au$^{77+}$ and uranium U$^{90+}$
ions. As seen from the results obtained, the photon emission pattern
appears to be sensitive to the symmetry and parity of the particular
excited state as well as to the higher multipole contributions to
the electron--photon interaction. The strongest non--dipole effects
have been identified for the $1s2s \, ^3S_1 \to 1 \, ^1S_0$
two--photon transition for which the 2E1 decay channel is forbidden
owing to symmetrization properties of the system. For the other two
transitions, $1s2s \, ^1S_0 \to 1 \, ^1S_0$ and $1s2p \, ^3P_0 \to 1
\, ^1S_0$, the higher multipoles of the radiation field typically
result in a 10--30 \% deviation of the photon--photon correlation
function from the (analytical) predictions obtained within the
dipole 2E1 approximation. This deviation becomes most apparent for
the parallel and back--to--back photon emission and may be observed
not only for high--$Z$ but also for medium--$Z$ ions.

\medskip

The second--order perturbation approach based on the independent
particle model, used in the present calculations, is appropriate for
the analysis of forthcoming experimental studies on the two--photon
transitions between the $^{2s+1}L_J$ excited and the ground states
of helium--like, heavy ions. Besides these spontaneous decays, whose
energies usually reach 100 keV, \textit{induced} $J=0 \to J=0 + 2
\gamma$ transitions between excited states are also likely to be
explored at the GSI ion storage ring \cite{ScS89}. Having energies
in the optical range (2--3 eV), these transitions may provide an
alternative and very promising tool for studying the parity
violation phenomena. Their theoretical analysis, however, requires a
more systematic treatment of the electron--electron interaction
effects. Based on the multi--configuration Dirac--Fock approach and
B--spline basis set method, investigations along this line are
currently underway and will be reported elsewhere.

%
%

\section*{Acknowledgements}

A.S. and F. F. acknowledge support from the Helmholtz Gemeinschaft
and GSI under the project VH--NG--421. S.F. acknowledges the support
by the DFG. This research was supported in part by FCT Project No.
POCTI/0303/2003 (Portugal), financed by the European Community Fund
FEDER and by the Ac\c{c}\~oes Integradas Luso-Alem\~as (Contract No.
A-19/09). A.V. and G.P. acknowledge support from the DFG and GSI.
Laboratoire Kastler Brossel is "Unit\'e Mixte de Recherche du CNRS,
de l' ENS et de l' UPMC No. 8552". This work is supported by
Helmholtz Alliance HA216/EMMI. PI acknowledge support from the PHC
program PESSOA 2009 number 20022VB.

%
%
%
%

\end{document}